\documentclass[superscriptaddress,letterpaper, aps, amssymb, preprint, amsmath, 12pt]{revtex4-1}

 \usepackage[justification=raggedright,singlelinecheck=false,labelfont=bf,labelsep=period]{caption,subcaption}

\usepackage{xr}

\usepackage{braket}

\usepackage{siunitx}

\usepackage{comment}

\usepackage{nameref}

\newcommand{\wmk}{Wm$^{-1}$K$^{-1}$}

\usepackage{scalerel}

\usepackage{tikz}

\usetikzlibrary{svg.path}
\definecolor{orcidlogocol}{HTML}{A6CE39}
\tikzset{
  orcidlogo/.pic={
    \fill[orcidlogocol] svg{M256,128c0,70.7-57.3,128-128,128C57.3,256,0,198.7,0,128C0,57.3,57.3,0,128,0C198.7,0,256,57.3,256,128z};
    \fill[white] svg{M86.3,186.2H70.9V79.1h15.4v48.4V186.2z}
                 svg{M108.9,79.1h41.6c39.6,0,57,28.3,57,53.6c0,27.5-21.5,53.6-56.8,53.6h-41.8V79.1z M124.3,172.4h24.5c34.9,0,42.9-26.5,42.9-39.7c0-21.5-13.7-39.7-43.7-39.7h-23.7V172.4z}
                 svg{M88.7,56.8c0,5.5-4.5,10.1-10.1,10.1c-5.6,0-10.1-4.6-10.1-10.1c0-5.6,4.5-10.1,10.1-10.1C84.2,46.7,88.7,51.3,88.7,56.8z};}}
\newcommand\orcidicon[1]{\href{https://orcid.org/#1}{\mbox{\scalerel*{
\begin{tikzpicture}[yscale=-1,transform shape]
\pic{orcidlogo};
\end{tikzpicture}
}{|}}}}

\usepackage{etoolbox}

\makeatletter
\pretocmd\frontmatter@thefootnote{\color{blue}}{}{}

\usepackage{dcolumn} 

\usepackage{titlesec}
\titlespacing*{\section}
{0pt}{4ex}{1.3ex}
\titlespacing*{\subsection}
{0pt}{4ex}{0ex}

\titleformat{\section}
  {\bfseries\MakeUppercase}{\thesection\raggedright}{1em}{}
  \titleformat{\subsection}
 {\bfseries}{\thesubsection\raggedright}{1em}{}

\usepackage{caption}
\usepackage{indentfirst}

\usepackage{xstring}
\usepackage{xspace}
\usepackage[utf8]{inputenc}
\usepackage{natbib}
\usepackage{gensymb}
\usepackage{mathtools}

\usepackage{float}
\usepackage{graphicx}

\usepackage{url}
\usepackage[utf8]{inputenc}
\usepackage[T1]{fontenc}
\usepackage{textcomp}
\usepackage{gensymb}
\usepackage[colorlinks,citecolor=red,urlcolor=blue,bookmarks=false,hypertexnames=true]{hyperref}
\usepackage{booktabs}
\usepackage{multirow}
\usepackage{textcomp}
\usepackage{graphicx}


\makeatletter
\let\@fnsymbol\@alph
\makeatother

\begin{document}
\title{A perspective on inelastic light scattering spectroscopy for probing transport of collective acoustic excitations}
\author{Hyemin Kim~\orcidicon{0009-0001-5560-7366}}
\thanks{Hyemin Kim and Hyungseok Kim contributed equally to this work.}
\affiliation{%
Department of Mechanical Engineering, Seoul National University, Seoul 08826, Republic of Korea}%
\author{Hyungseok Kim~\orcidicon{0009-0007-4572-567X}}
\thanks{Hyemin Kim and Hyungseok Kim contributed equally to this work.}
\affiliation{%
Department of Mechanical Engineering, Seoul National University, Seoul 08826, Republic of Korea}%

\author{Taeyong Kim~\orcidicon{0000-0003-2452-1065}}
 \email{tkkim@snu.ac.kr}
\affiliation{%
Department of Mechanical Engineering, Seoul National University, Seoul 08826, Republic of Korea}%
\affiliation{
Institute of Advanced Machines and Design, Seoul National University, Seoul 08826, Republic of Korea
}
\affiliation{%
Inter-University Semiconductor Research Center, Seoul National University, Seoul 08826, Republic of Korea}%

\begin{abstract}
Understanding and manipulating nanoscale energy transport and conversion processes are essential for diverse applications, ranging from thermoelectrics and energy harvesting to thermal management of microelectronics. While it has long been recognized that acoustic and thermal properties in condensed matters are primarily due to microscopic transport of phonons as quasiparticles, probing thermal acoustic excitations particularly at sub-THz remains a challenge primarily due to limitations in experimental techniques with spatiotemporal resolutions pertinent to probing them. Brillouin light scattering (BLS) and its variant, impulsive stimulated scattering (ISS), provide access to these thermal acoustic excitations, enabling measurement of quantities such as acoustic dispersions along with relaxation dynamics occurring in ultrasonic as well as hypersonic frequencies. In this perspective, we provide a brief overview of the operational principles of BLS and ISS, and highlight their applications in probing acoustic, thermal, and magnetic excitations in emerging and low-dimensional materials. We conclude by discussing current challenges and future opportunities for advanced material characterization using Brillouin light scattering spectroscopy techniques.
\end{abstract}

\maketitle

\section*{Introduction}
Nanoscale energy transport and conversion processes are essential for a wide range of applications in thermoelectrics, energy harvesting, and heat dissipation in modern devices \cite{Gang_2005_txtbk, Pop_2010_NanoR, Gaskins_2017_ElecSoc_highkdi_review, Scott_2018_APLmat_highkdi, Shakouri_2011_AnnuRev_thermoelectric1, Mukherjee_2022_JMCC_thermoelectric2, Zhang_2022_NatElec_thermoelectric3}. These processes are primarily facilitated by quasiparticles, among which phonons are dominant heat carriers in crystals~\cite{Cahill_2003_JAP_nanoheat, Cahill_2014_APR_nanoheat2}. On a microscopic scale, lattice heat conduction can be characterized by dispersion relations determined by crystal structure and relaxation dynamics that govern the transport properties of these heat-carrying quasiparticles. It has been demonstrated that long-wavelength, sub-THz thermal acoustic phonons contribute substantially to heat transport in bulk and nanostructured materials~\cite{Gang_2005_txtbk}, and that access to these modes is essential for understanding and tailoring their thermal and acoustic properties. Diverse scattering spectroscopy techniques have been developed and applied to probe the transport of thermal phonons in condensed matters, among which inelastic neutron scattering (INS) and inelastic X-ray scattering (IXS) are well-established methods for measuring vibrational dispersion relations and lifetimes \cite{Delaire_2011_Natmat_PbTe, Ferreira_2018_PRL, Masciovecchio_prb_1997, baldi_prl_2010,sette_science_1998}. However, both techniques face constraints arising from instrumental limitations and scattering kinematics. In IXS, its limited energy resolution restricts access to sub-THz vibrations~\cite{Wischnewski_1998_PRB, Baldi_2011_JNonCS}, and achieving high resolution requires high brilliance synchrotron sources, restricting its use to large-scale facilities~\cite{Baron_2016_intro}. INS, on the other hand, can provide enhanced energy resolution and is less invasive than IXS owing to high penetration depth of neutrons, which deposit relatively small energy in the sample, whereas X-rays typically have shallow penetration depth and can induce radiation damage~\cite{Butler_2022_book}. Nevertheless, since INS requires an ample sample volume (\textgreater mm$^3$~\cite{Piccoli_2007_INS}), it has limited sample coverage, especially for specimens that need to be synthesized as thin films, and kinematic constraints complicate measurements even when sufficient sample volume is available.

To probe thermal acoustic phonons beyond the spectral resolution of INS and IXS, Brillouin light scattering spectroscopy (BLS) has emerged as an established technique, complementing Raman spectroscopy which allows to probe zone-center optical polarizations~\cite{weber_2013_raman, Kargar_2021_natphot}. In principle, Brillouin scattering, also known as Brillouin–Mandelstam scattering, involves the scattering of an optical wave by an acoustic wave. The earliest theoretical works on Brillouin scattering were independently presented by Brillouin~\cite{brillouin1922} and Mandelstam \cite{mandelstam1926} in the early 20th century. The first experimental observation was reported shortly thereafter by Gross \cite{gross1930, gross1930_2}. The subsequent introduction of Fabry-Perot (FP) interferometers by Sandercock \cite{sandercock_1978_SSC} became a key component in modern BLS configurations, providing high contrast while maintaining moderate finesse to ensure sufficient transmittance of Brillouin signal for achieving high signal-to-noise ratio. BLS has since become widely accessible owing to their compact optical setup and automated FP interferometers, enabling routine measurements of acoustic excitations with sub-GHz spectral resolution in standard laboratory settings~\cite{hillebrands_revsciinst_1999, Bencivenga_revsciinst_2012}. Later, stimulated light scattering was developed as a means to efficiently amplify the scattered signal~\cite{Chiao_1964_PRL} whose intensity by the spontaneous process is known to be substantially weak~\cite{damzen_2003_stimulated, Boyd_2008_textbook}. The scattering process can be stimulated through the strong coupling between the incident light and the material, producing an exponentially amplified signal~\cite{Eggleton_NatPhoto_2019}. Building on this, by introducing a pump laser with a pulse duration less than period of oscillation (inverse of acoustic phonon frequencies), the pump pulses impulsively create spatially periodic and time-dependent excitations in the material \cite{Yan_1985_JCP_ISS1, Yan_1987_JCP_ISS2}. The following modulations in the optical properties of the material can be measured using a probe beam. This method, termed impulsive stimulated scattering, was first demonstrated by Boersch and Eichler in 1967~\cite{Boersch_1967_firstISBS}. We will refer to this process as impulsive stimulated scattering (ISS) when the scattering occurs in a stimulated manner by an impulsive light source, whereas we will specify the term Brillouin light scattering (BLS) when addressing spontaneous or general Brillouin scattering.

Due to its non-contact and non-destructive nature, BLS has been extensively employed for characterizing photoelastic properties of crystalline and glassy materials across a range of phases~\cite{letoublon2016elastic_BLSapplication1, mao2018solids, sandercock2007some, ko2011precursor, benedek_ieee_1965_liquid}. In theory, spontaneous Brillouin scattering involves no optical absorption, therefore impacts of heating in BLS are typically insufficient to induce sample damage~\cite{Nikolic_2019_biomedoptexp}. Moreover, unlike conventional static methods based on finite-deformation stress-strain measurements, which typically yield a few elastic constants under stringent orientation conditions~\cite{Koski_2013_natmat, Hearmon_1946_revmodphys}, BLS enables evaluation of the full elastic tensor in anisotropic materials and oriented fibers~\cite{Koski_2013_natmat, Elsayad_2020_cellulose}.  In addition to elastic measurements, BLS has emerged as a routine tool for probing properties of quasiparticles such as phonons and magnons. In the case of phonons, BLS enables access to low-frequency phonon dispersion, especially in nanostructured solids. For instance, BLS has been used to observe confined phonon modes in GaAs nanowires~\cite{Kargar_2016_Natcom_BLSGaAs}, and to resolve gigahertz phonon dispersion relations in patterned holey silicon membranes, thereby providing direct experimental evidence of omnidirectional vibrational band gaps in phononic crystals~\cite{Florez_2022_NatNanotech}. Beyond acoustic excitations, BLS can also probe magnons, which are of interest for spintronics applications \cite{Demidov_2009_prl, Vogt_2014_natcomm, Nembach_natphys_2015, Holanda_2018_natphys, Demidov_2016_natcomm}. BLS has been employed to spatially map magnon propagation in magnonic waveguides~\cite{demidov_apl_2007, vogt_apl_2012, Demidov_2016_natcomm}, to quantify interfacial Dzyaloshinskii-Moriya interaction strength in metal multilayers~\cite{Nembach_natphys_2015, cho_natcommun_2015} and, notably, to demonstrate magnon Bose-Einstein condensate at room temperature~\cite{Demokritov_2006_Nat_BoseEinstein, Borisenko_2020_natcomm}. However, despite these capabilities, BLS suffers from an inherently weak scattering signal, resulting in long acquisition times and low signal-to-noise ratio (SNR)~\cite{Kabakova_natrev_2024, li_photonres_2024, Bilenca_2024_jphysphot}.

ISS can access dynamical responses with temporal resolution below nanosecond and has tunable accessible length scales, providing enhanced signal intensity with an additional laser beam, though optical wave mixing geometry is required~\cite{Choudhry_2021_JAP}. Here, ISS can be categorized into impulsive stimulated Brillouin scattering (ISBS), which arises from electrostriction in transparent materials, and impulsive stimulated thermal scattering (ISTS), which occurs in optically absorbing media. While ISBS has been traditionally used to measure viscoelastic properties of liquids~\cite{yan_jchemphys_1988, Silence_1992_jchemphys, Le_2024_jphysphot}, recent development of ISBS microscopy has shifted its focus toward elastic characterizations of biological specimens~\cite{Ballmann_2017_optica, krug2019impulsive, li_photonres_2024}. Owing to enhanced intensity of Brillouin signal generated through impulsive stimulated processes, ISBS offers improved spectral resolution, shorter acquisition times, and higher SNR compared to BLS~\cite{Prevedel_2019_NatMethods_BLSmicroscopy}. However, it is more invasive due to the increased optical power exposure, which can lead to photodamage in sensitive and fragile specimens~\cite{Chow_2023_optexp}. This has motivated ongoing efforts to mitigate laser-induced damage while enhancing SNR~\cite{li_photonres_2024}. Another active area of research in ISBS focuses on pushing the limit of spatial resolution, as its current resolution remains limited to several tens to hundreds of micrometers~\cite{Zhang_2021_natprot}.

ISTS has been widely applied to study thermal and elastic properties of thin films, suspended membranes, and bulk materials \cite{Johnson_2012_JAP, Johnson_2013_PRL_TGsignals, cuffe2015reconstructing, li2021remarkably}, as well as highly anisotropic materials and soft matters \cite{rogers2000optical, hecksher2017toward, khanolkar_apl_2015}. A key feature of impulsive stimulated Brillouin scattering is its tunable accessible length, which is pertinent to probing transport processes across different spatio-temporal scales. This extends measurements of heat transport beyond the diffusive regime, allowing experimental reconstruction of vibrational mean free paths (MFPs) spectra in lattices with various phases by resolving ballistic heat transport~\cite{Johnson_2013_PRL_TGsignals, cuffe2015reconstructing, Andrew_2019_PNAS_DrawRatio,Kim_2021_PRM_aSi_TG}. Additionally, beyond mean free path spectroscopy, it has been applied to detect second sound in graphite at markedly high temperatures by accessing phonon hydrodynamic transport regime~\cite{Huberman_2019_Science, ZDing_2022_NatComm}. Another variant of ISS is spin grating spectroscopy, which has been applied to probe electronic spin dynamics and valley polarization dynamics: discovery of spin-Coulomb drag in GaAs quantum wells; observation of the persistent spin helix in semiconductor quantum wells; measurement of valley depolarization dynamics in low-dimensional materials \cite{weber_nature_2005, koralek2009emergence, Mahmood_2018_NanoLetters_ISBSapp2}.

The interpretation of BLS and ISS measurements require subsequent signal processing and data analysis. For post-processing in BLS, the measured BLS spectrum must be corrected for instrumental broadening arising from the laser linewidth and the finite finesse of FP interferometers~\cite{Hwa_joptsoc_1989}. The observed Brillouin linewidth therefore represents a convolution of the intrinsic phonon linewidth with the instrument response function~\cite{Bouvet_natphot_2025}. The intrinsic linewidth, hence the phonon lifetime, is obtained by deconvolving this response through analytical or numerical fitting approaches ~\cite{Hwa_joptsoc_1989, Vanderwal_1981_optcommun}. In ISBS, the impulsive excitation of the medium by crossed pump pulses launches damped standing-wave oscillations~\cite{Yan_1987_JCP_ISS2}. Traditionally, the resulting signal is analyzed by nonlinear least-squares fitting to a damped sinusoidal function to extract the acoustic frequency and attenuation rate. Subsequently, viscoelastic properties are determined using relaxation models, such as Debye or Kohlrausch-Williams-Watts (KWW) models~\cite{Yan_1987_JCP_ISS2, Duggal_1991_jchemphys}. While ISBS and ISTS measurements share similar geometry, ISTS involves temperature rise due to optical absorption, whereas ISBS is excited by electrostrictive forces~\cite{Torchinsky_JCP_2009}. In ISTS, signals from the photoexcitation process by the impulsive laser beam primarily comprise electronic and thermal decay responses of the absorbing material. Rather than directly deconvoluting the signal, the fitting function is convolved with the impulse response, which contains the instrument's response~\cite{Andrew_2019_PNAS_DrawRatio, Kim_Natcomm_2022}. The electronic and thermal relaxation dynamics are then obtained by extracting the decay time constants from the fits of the measured signals, and further analysis of phonon dynamics can be conducted based on grating period dependency.



In this mini-review, we provide a brief overview of working principles and instrumentations of modern BLS and its variant, ISS. We then highlight recent studies that employ BLS and ISS to investigate various modes of nanoscale energy transport, including acoustic, thermal, and magnetic excitations. The purpose of this review is to provide an accessible introduction to these photonic techniques, with a perspective on their evolving applications. This manuscript is organized as follows: we first introduce brief operational principles of BLS, and survey its contemporary applications; next, we then present ISS in the same order. This review aims to provide the scientific community with our perspective on current trends and future opportunities for advancing Brillouin light scattering spectroscopy as a versatile tool for characterizing energy transport and conversion across emerging materials and nanoscale systems, extending its scope beyond conventional applications.

\section{Working principles and instrumentation of BLS}
\begin{figure}[hbt!]
{\includegraphics[width=16cm,height=12cm,keepaspectratio]{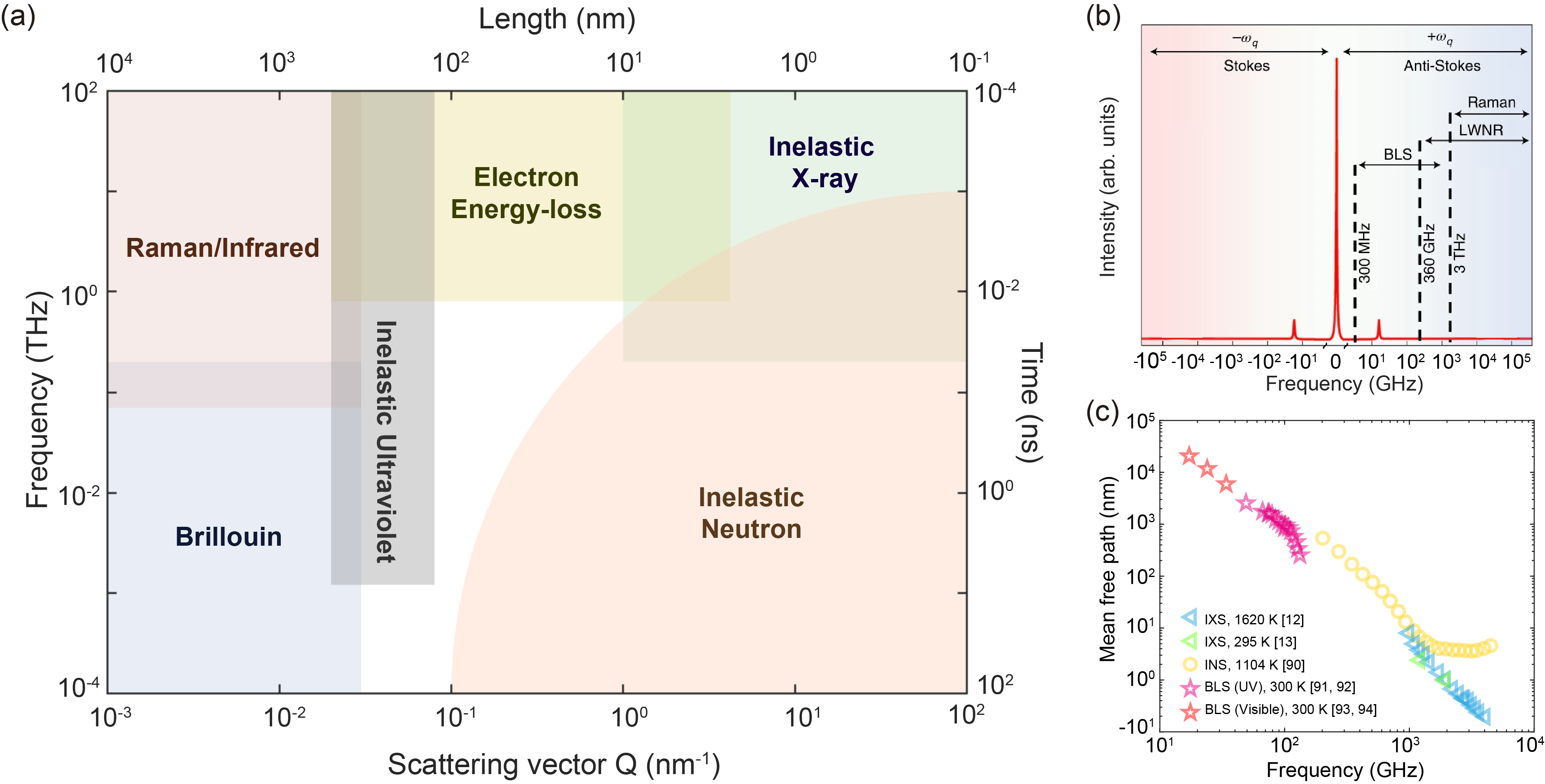}}
\caption{(a) Map showing the accessible frequency and momentum transfer ranges of different inelastic scattering spectroscopy techniques. Shaded regions indicate typical accessible windows for each technique, and the Brillouin scattering shows the appropriate range to probe excitations with relatively low frequency and momentum. Adapted from Ref.~\cite{Bencivenga_APX_2023} with permission under CC BY license. 
(b) Representative spectra for Brillouin scattering, showing inelastic Stokes (downshifted) and anti-Stokes (upshifted) peaks displaced from the central elastic Rayleigh scattering peak by Brillouin shift. The shifted peaks exhibit relatively low intensity compared to Rayleigh scattering, and the spectral separation is determined by the wavelength and angle of incident light, together with the refractive index and acoustic velocity of a material. The dotted lines represent the measurement range of BLS, Raman, and low-wavenumber Raman (LWNR) spectroscopy. Reprinted with permission from Karger, F. \textit{et al.}, \textit{Nature Photon.} \textbf{15}, 720-731 (2021). Copyright 2021 Springer Nature. (c) Measured mean free paths versus frequency for thermal acoustic vibrations in vitreous silica from IXS (triangles, Ref.~\cite{baldi_prl_2010, Masciovecchio_prb_1997}), INS (open circles, Ref.~\cite{Wischnewski_prb_1998}), BLS (5-pointed stars, Ref.~\cite{Masciovecchi_prl_2006, Benassi_prb_2005, Vacher_prb_2006, Levelut_prb_2006}).}
\label{fig:figure1}
\end{figure}

Unlike other inelastic scattering spectroscopies such as inelastic X-ray scattering, which are limited to resolve excitations with frequencies above a few THz, Brillouin scattering offers access to acoustic excitations with frequencies ranging from sub-gigahertz up to several hundreds gigahertz, as shown in Fig.~\ref{fig:figure1}(a). In this section, we provide a brief introduction on the working principle of Brillouin scattering spectroscopy techniques, and outline modern experimental configurations that enable such spectral resolution in these photonic approaches.

\subsection{Stokes and anti-Stokes scattering}
Typical BLS spectra exhibit several intensity peaks as shown in Fig.~\ref{fig:figure1}(b): a central peak occurring at zero frequency is due to elastic Rayleigh scattering, along with other frequency-shifted inelastic peaks. The downshifted peak is due to Brillouin scattering in which light is scattered by interaction with acoustic waves, namely phonons, in the medium. In this process, called the Stokes process, phonons are created during the light-matter interaction, resulting in a lower frequency of the scattered wave. In contrast, the upshifted peak at higher frequency corresponds to the anti-Stokes scattering where phonons are annihilated during the interaction, causing a frequency increase in the scattered light. For both Stokes and anti-Stokes Brillouin peaks, the frequency shifts relative to elastic peak correspond to the frequency of excitations (e.g., phonons) with wavevectors following the dispersion relation, while the widths of the inelastic peaks reflect the phonon lifetimes. The phonon wavevector can be calculated from the energy and momentum conservation of Brillouin scattering and the phonon lifetime can be determined from Lorentzian fitting of the Brillouin peaks.

During inelastic light scattering processes, the incident and scattered light with wavevectors $\mathbf{k}$ and $\mathbf{k'}$, respectively, should satisfy the conservation of momentum $\mathbf{k'} = \mathbf{k} \pm \mathbf{q}$, where $\mathbf{q}$ is the momentum transfer, identical to the wavevector of an acoustic wave with frequency $\nu$. Here, the momentum transfer is substantially smaller compared to the incident and scattered light wavevectors, and thus can be considered negligible. Then, the momentum transfer becomes $\left|\mathbf{q}\right|=2\left| \mathbf{k} \right|\sin{\left ({\theta}/{2}\right)}$, where $\theta$ is the scattering angle. Finally, following the dispersion relation $\nu = {\left| \mathbf{q} \right| v_s}/{2\pi}$, the frequency shift $\nu$ of the measured lines, which is identical to the frequency of the acoustic wave with wavevector $\mathbf{q}$, is expressed as~\cite{Boyd_2008_textbook}
\begin{equation}
\nu = \frac{2nv_s}{\lambda_i} \sin \left ({\theta}/{2} \right)
\label{eq:freqshift}
\end{equation}
where $\lambda_i$ is the wavelength of the incident optical wave, $n$ is the refractive index, and $v_s$ is the acoustic velocity. As can be seen in Eq.~\ref{eq:freqshift}, it is apparent that the magnitude of the frequency shift is maximum (minimum) at scattering angle of 180\degree (0\degree). This indicates that backscattering geometry ($\theta = 180\degree$) provides access to the maximum detectable phonon wavevector along with the corresponding frequency, while spectrum of forward scattering will be dominated by Rayleigh scattering due to the smaller momentum transfer. The frequency shift in Brillouin scattering can also be interpreted as a Doppler effect between light and the acoustic wave. More precisely, the redshift observed in Stokes scattering is analogous to light scattering from a receding acoustic wave in the material, while the blueshift of the scattered wave in anti-Stokes scattering corresponds to scattering from an approaching acoustic wave \cite{Boyd_2008_textbook, Benedek_1966_PhysRev}.

An additional physical property can be extracted from analyzing the peak widths of BLS spectra. As shown in the peaks of Fig.~\ref{fig:figure1}(b), shifted peaks manifest a finite width due to acoustic attenuation. The linewidths of the peaks are directly related to the phonon lifetimes, where phonon lifetime $\tau_p$ can be determined using full width at half maximum (FWHM) of the Brillouin peak using $\Delta \nu = 1/2\pi\,\tau_p = \Gamma {\lvert \mathbf{q}\rvert}^2 / 2 \pi$. Here, $\Delta \nu$ is the FWHM and $\Gamma$ is the attenuation coefficient \cite{berne_2000_text}. Noting that the momentum transfer is $\lvert\mathbf{q}\rvert=2\lvert \mathbf{k}\rvert\sin(\theta/2)$, lifetime is determined as follows~\cite{Boyd_2008_textbook}.
\begin{equation}
\tau_{p} = \frac{1}{4\,\Gamma\,\lvert \mathbf{k}\rvert^{2}\,\sin^{2}\!\bigl({\theta}/{2}\bigr)} = \frac{{\lambda_i}^{2}}{16\,n^{2}\,\pi^2\,\Gamma\,\sin^{2}\!\bigl({\theta}/{2}\bigr)}
\label{eq:lifetime}
\end{equation}

From Eq.~\ref{eq:lifetime}, the width of Brillouin peaks directly reflects the lifetimes of phonons which participate in the scattering process, and provides insight into the transport properties of low-energy phonons in the material along with the physical properties such as elastic constants. For example, in liquids, sub-ns vibrational lifetimes at a few gigahertz were measured in toluene and methylene chloride, and the results reflect the viscosity coefficients of such media \cite{shapiro_ieee_1966}. The vibrational lifetimes in several glasses were estimated to be in the range of 2-4 ns at hypersonic frequencies, falling between the values of organic liquids and crystalline quartz, where the latter exhibits a phonon lifetime of $\sim5\,\mathrm{ns}$~\cite{Kaiser_PRB_1970}. Given that lifetimes of acoustic vibrations in glasses lie between those of crystals and liquids, it indicates that vibrational lifetimes depend on the underlying lattice structure and phase of a material. Furthermore, the mean free paths of thermal acoustic excitations can be obtained from the measured lifetimes and velocity determined by BLS. Figure~\ref{fig:figure1}(c) illustrates the measured mean free paths versus frequency for vitreous silica obtained from BLS, INS, and IXS, highlighting how BLS bridges the low-frequency regime inaccessible using conventional inelastic scattering spectroscopy techniques.

\subsection{Spontaneous Brillouin scattering spectroscopy}
In principle, Stokes and anti-Stokes scattering of spontaneous Brillouin scattering are driven by modulations in the dielectric constant due to thermal fluctuations and zero-point energy~\cite{damzen_2003_stimulated, Boyd_2008_textbook}. Such variations in the optical properties lead to inelastic scattering of the incident light, and two types of excitations are involved depending on the optical characteristics of the materials. In weakly absorbing materials such as optically transparent or semitransparent media, where their optical penetration depth is sufficiently larger than the wavelength of incident light, the BLS signal primarily arises from bulk phonons via opto-elastic coupling~\cite{Bottani_APX_2018}. In opaque or strongly absorbing materials, if penetration depth is limited to the surface due to high optical extinction of the medium, scattering predominantly occurs by surface phonons~\cite{briggs_text_2013}. In either case, the magnitude of the measured phonon wavevector depends on the incident angle and the wavelength of the excitation laser as described in Eq.~\ref{eq:freqshift}.

\begin{figure}
{\includegraphics[width=16cm]{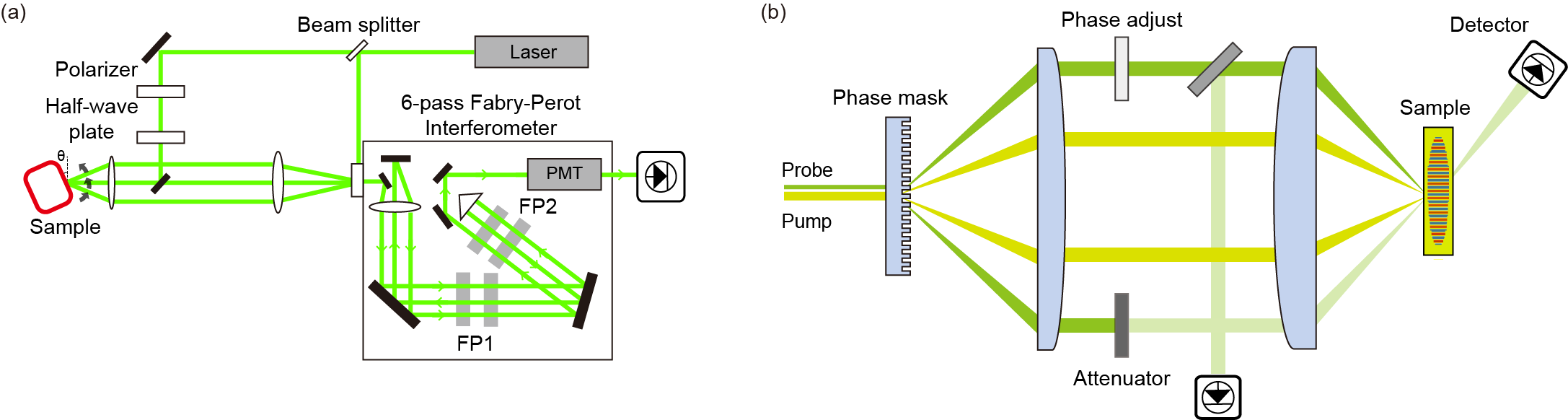}}
\caption{Typical experimental setup for (a) spontaneous Brillouin light scattering spectroscopy, adapted with permission from DOI:10.1088/0022-3727/45/27/275302, \textit{J. Phys. D: Appl. Phys.}, \textbf{45}, 275302. Copyright 2012 IOP Publishing Ltd \cite{BLS_setup}, and (b) impulsive stimulated scattering spectroscopy.}
\label{fig:expt_setup}
\end{figure}

A schematic of a representative experimental setup used for backscattering BLS is shown in Fig.~\ref{fig:expt_setup}(a). A laser beam is initially directed toward a beam splitter, where the reflected light is used to define a zero-frequency position for calibrating 3FP interferometers~\cite{Bencivenga_revsciinst_2012}. The transmitted beam is refined by a polarizer, then its polarization is adjusted by a half-wave plate to enhance sensitivity to the quasiparticle mode of interest~\cite{heiman_prb_1979}. The sample is typically mounted on a motorized rotation stage to vary the incidence angle of the laser~\cite{link_japhys_2006, pang_adphr_2022}, which provides access to different phonon wavevectors in opaque samples, whereas in bulk transparent materials, phonon wavevector is determined by the refractive index of the medium as $\lvert \mathbf{q}\rvert=4\pi n/\lambda_i$ in backscattering geometry~\cite{mutti_adv_1995}. The scattered light is then focused into FP interferometers, which provide a high spectral contrast to isolate weak Brillouin peaks from strong elastic Rayleigh line while suppressing parasitic transmission peaks and maintaining high effective transmittance via tandem design of interferometers~\cite{sandercock_book_1982, Blachowicz_1996_revsciinst, Dil_1981_applopt, Scarponi_prx_2017}. The output is detected by a photomultiplier tube in single-photon counting mode to capture low-intensity scattered light~\cite{Witschas_2010_applopt}. With sub-GHz spectral resolution, modern BLS configuration can detect weak optical signals, which allows access to hypersonic phonons and thermal magnons~\cite{sandercock_book_1982, Bencivenga_revsciinst_2012, Birt_2013_apl}, even in thin film with a few Angstroms~\cite{hillebrands_prb_1987}. Another advantage of such configuration is automation via self-aligning FP interferometers~\cite{hillebrands_revsciinst_1999}, allowing versatility in typical lab settings. Adding an objective lens with a high numerical aperture to achieve small laser spot and integrating an automated 3D scanning stage yield micro-focused BLS, or $\mu$-BLS, which has been used for spatiotemporal imaging of surface phonon and spin waves~\cite{geilen_apl_2020,yang_pra_2025,camley_solphys_2012}.

\begin{figure}
{\includegraphics[width=16cm,height=12cm,keepaspectratio]{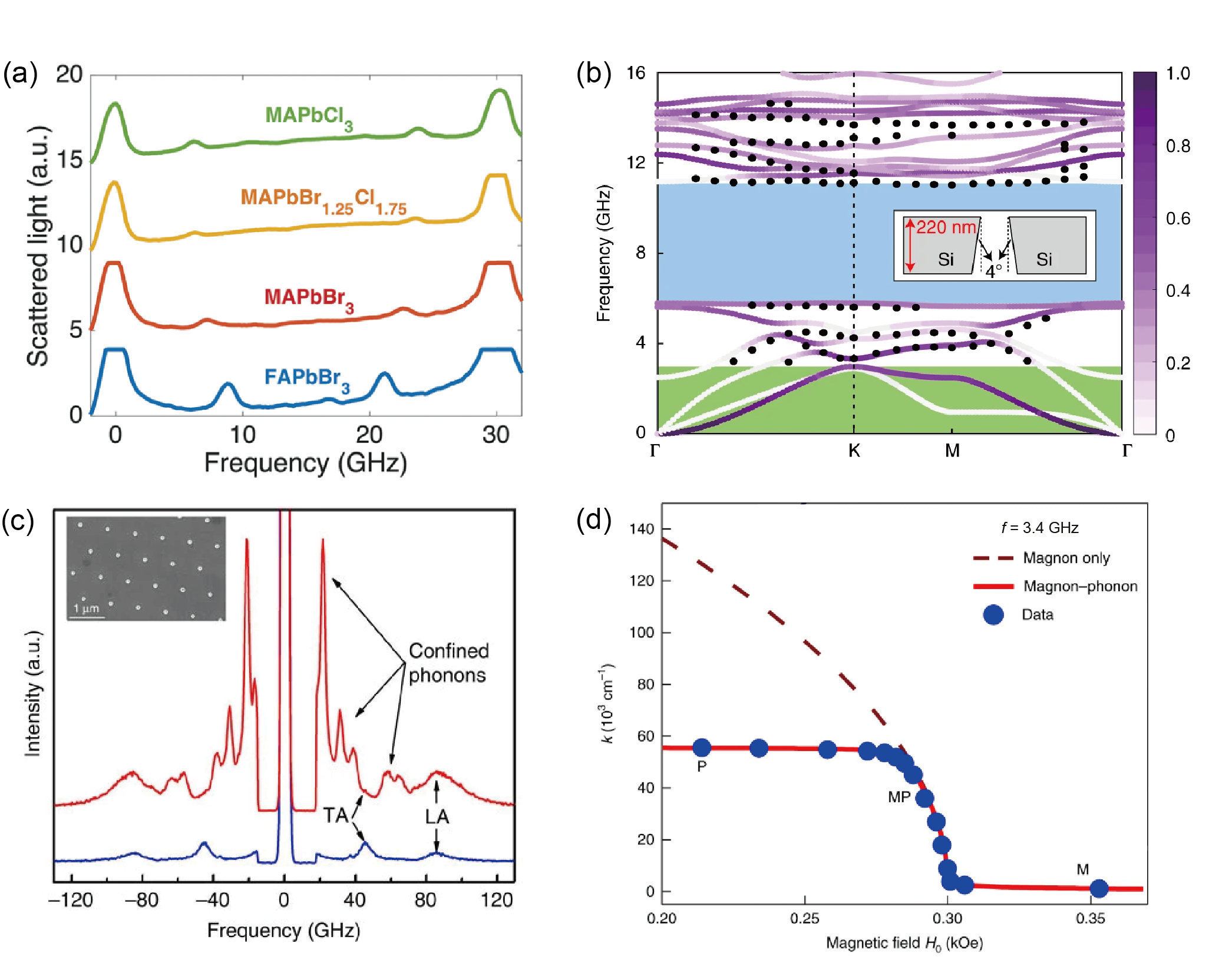}}
\caption{(a) Comparison of Brillouin spectra for hybrid halide perovskites with different cations and halides compositions, showing larger Brillouin shifts with increasing MA and Cl content. Reproduced from Ref.~\cite{kabakova2018effect} with permission from the Royal Society of Chemistry. (b) Phonon dispersion relation in a holey Si membrane obtained from BLS compared with simulations, showing an omnidirectional band gap~\cite{Florez_2022_NatNanotech}. Reprinted with permission from Florez, O. \textit{et al.}, \textit{Nature Nanotechnol.} \textbf{17}, 947-951 (2022). Copyright 2022 Springer Nature. (c) Measured Brillouin spectra of GaAs nanowires array (red) compared to bulk GaAs substrate (blue), showing additional low-frequency peaks corresponding to confined acoustic phonon modes. Reproduced from Ref.~\cite{Kargar_2016_Natcom_BLSGaAs}, licensed under a Creative Commons Attribution 4.0 International License. (d) Wavenumber-resolved BLS signal from a YIG thin film, showing field positions of maximum BLS amplitude, which corresponds to hybridized magnon-phonon modes~\cite{Holanda_2018_natphys}. Adapted with permission from J. Holanda \textit{et al.}, \textit{Nature Phys.} \textbf{14}, 500-506 (2018). Copyright 2018 Springer Nature.}
\label{fig:BLS_app_fig3}
\end{figure}

\section{Applications of BLS}
Conventionally, one of the earliest and continuing applications of BLS is characterizing elastic properties such as sound velocities and elastic moduli of materials~\cite{grimsditch_prb_1975, vacher_prb_1972, whitfield1976elastic, yamaguchi_jap_1999}, particularly bulk crystals and minerals under high pressures up to $\sim64$ GPa and temperatures above 1800 K~\cite{whitfield1976elastic, zha_pnas_2000, jiang_physearth_2009, raetz_prb_2019,Kurnosov_2024_physchem}, where traditional mechanical testing is challenging to apply and only feasible up to a few GPa due to large sample size constraints and difficulties in ensuring sufficient transducer-sample adhesion~\cite{Brazhkin_jexptheo_2001,  grimsditch1986brillouin, polian_prb_1982, zouboulis_prb_1998}. 

More recently, Brillouin spectroscopy has been employed to characterize elastic properties of novel materials, including soft lattices of hybrid halide perovskites~\cite{letoublon2016elastic_BLSapplication1, kabakova2018effect}; as shown in variations in Brillouin shifts from hybrid perovskites with different ionic compositions in Fig.~\ref{fig:BLS_app_fig3}(a), it was demonstrated that adjusting cation and halide composition can stiffen their lattices~\cite{kabakova2018effect}. BLS has also been applied to complex crystals~\cite{stevens_jchemphys_2005, tan_prl_2012, sanchez-valle_jap_2005, dirras_matscieng_2016, pang_adphr_2022}, where a recent study used Brillouin scattering to obtain an elastic tensor of a topological insulator (Sb$_2$Te$_3$), resolving its out-of-plane elastic constant $c_{33}$ by measuring velocity of surface acoustic modes~\cite{Baranowski_2025_scirep}. Likewise, BLS has been applied to 2D layered crystalline materials and polycrystalline thin films to measure elastic anisotropy~\cite{faurie_actmat_2010, jimenez-rioboo_apl_2018, reed_jap_2022, reed_acsnano_2024}. For instance, in few-layer MoSe$_2$ membranes, BLS was used to detect a $\sim30\%$ reduction in their in-plane stiffness compared to bulk, confirming elastic softening as the number of layers decreases~\cite{Babacic_2021_advmat}. Extending its scope, theoretical work has proposed that Brillouin scattering could provide access to Majorana fermions and other low-energy boundary modes arising in quantum spin liquids, in particular $\alpha-\mathrm{RuCl}_3$ 2D honeycomb structure, where these excitations are inaccessible using standard experimental probes such as angle-resolved photoemission spectroscopy due to their chargeless nature~\cite{Perreault_PRB_2016_MFermions}.

In addition to elastic property measurements, BLS has also become an indispensable method for probing dispersion relations and lifetimes of gigahertz vibrational modes in nanostructures and phononic crystals~\cite{balandin2012phononics, Zhang_2012_apl, Hou_2014_apl, Mielcarek_2012_pss, Graczykowski_2012_prb, nataj_2023_natcomm, Huang_2020_nanotech, Florez_2022_NatNanotech, wright_apl_2024}. In periodic nanostructured solids, BLS can directly observe phenomena such as Bragg band gaps, zone folding of acoustic phonon branches, and localized resonant modes \cite{Hou_2014_apl, Mielcarek_2012_pss, Yudistira_2016_prb}. It is generally accepted that in surface acoustic waves, periodically arranged scatterers (holey structure) cause phonon band folding and Bragg gaps~\cite{Florez_2022_NatNanotech},~as shown in Fig.~\ref{fig:BLS_app_fig3}(b), while periodic arrays of local resonator (pillar-based) structures introduce additional low-frequency gaps~\cite{Graczykowski_PRB_2015, Sledzinska_2020_advfuncmat}. Another interesting phenomenon in nanostructures is phonon confinement induced by boundary conditions, leading to modified phonon dispersion that can affect thermal conductivity~\cite{balandin_prb_1998, kargar_apl_2015, cuffe_nanolett_2012, neogi_acsnano_2015, reed_nanolett_2019}. As demonstrated in Fig.~\ref{fig:BLS_app_fig3}(c), BLS was employed to observe multiple acoustic phonon subbands in free-standing GaAs nanowires, confirming the existence of confined phonon modes in structures with length scales on the order of hundred nanometer~\cite{Kargar_2016_Natcom_BLSGaAs}.

Another class of quasiparticles accessible by BLS are spin waves, also known as magnons. In magnetic thin films and multilayer heterostructures, BLS is one of the established techniques to measure dispersion relations and damping of spin waves~\cite{sandercock2007some, Sandercock_1978_ieee, Grimsditch_1979_prl, an_prb_2014, stashkevich_prb_2015, ma_prl_2018}. BLS has been used to spatially map the intensity of the Brillouin peak due to magnons under external magnetic field, thereby measuring the decay length of spin waves in thin metal strips~\cite{vogt_apl_2009,Demidov_2009_prl, demidov_ieee_2015}. Magnetoelastic waves are another area of interest, in which BLS has been employed to probe magnon-phonon interactions~\cite{agrawal_prl_2013, zhao_physrevappl_2021, holanda_apl_2021, kunz_apl_2024}, demonstrating magnon-phonon conversion that yields a plateau in the field position since phonon modes are independent of the external magnetic field, as shown in Fig.~\ref{fig:BLS_app_fig3}(d), and further revealing that spin angular momentum can be transferred from magnons to phonons~\cite{Holanda_2018_natphys}. Analogous to phonons, magnon confinement effects were observed using BLS and $\mu$-BLS in metallic thin films~\cite{Demokritove_2001_physrep, wang_prx_2023, demidov_apl_2008, Kargar_2020_jmagmag}, highlighting possible strategies to manipulate magnon transport for spintronics applications.

\section{Working principles and instrumentation of ISS}
Impulsive stimulated light scattering is a method that probes light scattering using impulsive pump-probe setups to detect the transport and relaxation processes of energy carriers such as phonons and electrons  \cite{Eichler_2013_textbook}. The scattering can be stimulated when the acoustic wave in a material is induced through the excitation process with the incident light. 

The excitation process through the interaction of the laser and acoustic waves in a material is explained by two primary mechanisms: electrostriction and optical absorption. First, electrostriction is a mechanism in which mechanical stress in materials varies under an applied electric field, thereby enabling the stimulated scattering in non-absorptive materials \cite{Kroll_JAP_1965, Beugnot_PRB_2012}. Second, photothermal excitation can occur in absorbing materials \cite{Herman_PRL_1967, Kaiser_PRB_1970}, where optical absorption leads to localized heating and subsequent thermal expansion \cite{Boyd_2008_textbook}. In general, the stimulated stress in a sample is generated by a combination of both electrostriction and absorptive effects, with the relative contribution of each mechanism depending on the properties of the material.

In non-absorptive media, Brillouin scattering process can be stimulated through the process of electrostriction. ISBS is an excitation mechanism that utilizes electrostriction to generate dynamic gratings.
The interference pattern modulates dielectric properties at regions of high electric field intensity, inducing a periodic density grating through electrostriction~\cite{Kroll_JAP_1965, Beugnot_PRB_2012}. This density wave propagates at the material's sound velocity, with stress profiles that can be described by an impulse response. The generated acoustic waves exhibit dependence on the material's intrinsic viscoelastic properties through their velocity and attenuation characteristics~\cite{Torchinsky_JCP_2009}.

On the other hand, for absorptive media, ISTS can measure acoustic properties of materials from their response to impulsive photothermal excitation \cite{Rogers_2002_textbook_ISTS}. Incident optical beams excite the electrons, which, subsequently thermalize through recombination and scattering processes with the lattice, imparting their energy to phonons. This non-radiative recombination process of excited electrons dissipates energy as heat through phonons and includes the relaxation process after the Auger and Shockley–Read–Hall (trap-assisted) recombination. As a result, following the intensity grating of incident beams, a spatially periodic transient temperature profile is formed \cite{Choudhry_2021_JAP}. The resulting thermal expansion produces a subsequent stress, which launches counter-propagating coherent acoustic waves \cite{kim_aplmat_2017}.

As an experimental method for measuring scattered light, ISS can be utilized in the form of transient grating spectroscopy (TGS). TGS is a simplified variant of four-wave mixing that employs two temporally coincident, coherent pump beams crossed at an angle to generate a spatially modulated excitation field in a sample. The interference between these two coherently superimposed optical pump waves creates a spatially periodic interference pattern, which can then be measured by probe beams~\cite{Johnson_2012_JAP, vegaflick_revsciins_2015, Dennett_JAP_2018, Choudhry_2021_JAP}. Through variations in the real and imaginary refractive indices $n$ and $k$, the interference pattern generates phase and amplitude transient gratings, respectively, and the detector observes the probe beam diffracted from these gratings.

Contemporary TGS configuration that involves the phase mask and implements heterodyne detection is presented in Fig.~\ref{fig:expt_setup}(b). Heterodyne detection is a phase-sensitive detection technique that mixes the diffracted beam with a reference beam~\cite{Maznev_1998_OptLett, Goodno_Opt_1998, Torre_PRE_2001, Taschin_PRE_2006}. It offers advantages in achieving a high signal-to-noise ratio by suppressing unwanted noise components, while not saturating the detector. The first-order diffracted beams from the phase mask are selected as probe beams. One beam then passes through an optical phase adjust and acts as the signal beam to detect material properties, while the other beam is attenuated and employed as the reference beam in heterodyne detection. After Bragg diffraction of the signal beam at the sample, the diffracted signal beam is now superimposed onto the reflected or transmitted reference beam, depending on whether reflection or transmission geometry is used. For a thin sample under the transmission geometry, the intensity of the first-order diffracted probe beam under direct (homodyne) detection, assuming small refractive index modulation, is given by~\cite{Eichler_2013_textbook}
\begin{equation}
I_{\mathrm{df},t}
  \;=\;
  I_{\text{sig,td}}
  \,
  \Bigl[ (\Delta n)^{2} + (\Delta k)^{2} \Bigr]
\end{equation}
where $\Delta n$ and $\Delta k$ are the amplitude and phase grating signal caused by the real and imaginary parts of the refractive index respectively, and $I_{\text{sig,td}}$ is the intensity of signal beam transmitted through the specimen in the presence of the thermal grating. In contrast, the time-dependent signal intensity from the heterodyne detection is
\begin{equation}
I_{\text{hd},t} = I_{\text{sig,td}} \, 2t_r  \Bigl[ \Delta n \sin \Delta \phi - \Delta k \cos \Delta \phi \Bigr] 
\label{eq:heterodyne}
\end{equation}
and $t_{r}$ is the signal-to-reference transmission coefficient. Equation~\ref{eq:heterodyne} indicates that several advantages can be obtained through the implementation of heterodyne detection. First, scattered signal magnitude can be increased by the attenuation factor ($t_{r}$), thereby achieving a high signal-to-noise ratio. Second, contributions from the amplitude and phase gratings can be decoupled by adjusting the relative phase $\Delta\phi$ between the signal and reference beams to be multiples of $\pi/2$. More precisely, through this phase adjustment, heterodyne detection produces a signal that varies with the material response parameters $\Delta n$ or $\Delta k$ and enables separation of the amplitude grating contribution from $\Delta k$ and the phase grating contribution from $\Delta n$~\cite{Choudhry_2021_JAP}. Additionally, subtracting experimental results between different $\Delta\phi$ can filter out unwanted signal components while remaining signal is doubled. Using this transient thermal grating approach, TGS provides information about thermoelastic properties of materials, such as thermal conductivity and elastic constants, through thermoelastic response measurements across various research fields, with specific applications detailed in a later section.

\begin{figure}
\centering
{\includegraphics[width=14.5cm,height=10.5cm,keepaspectratio]{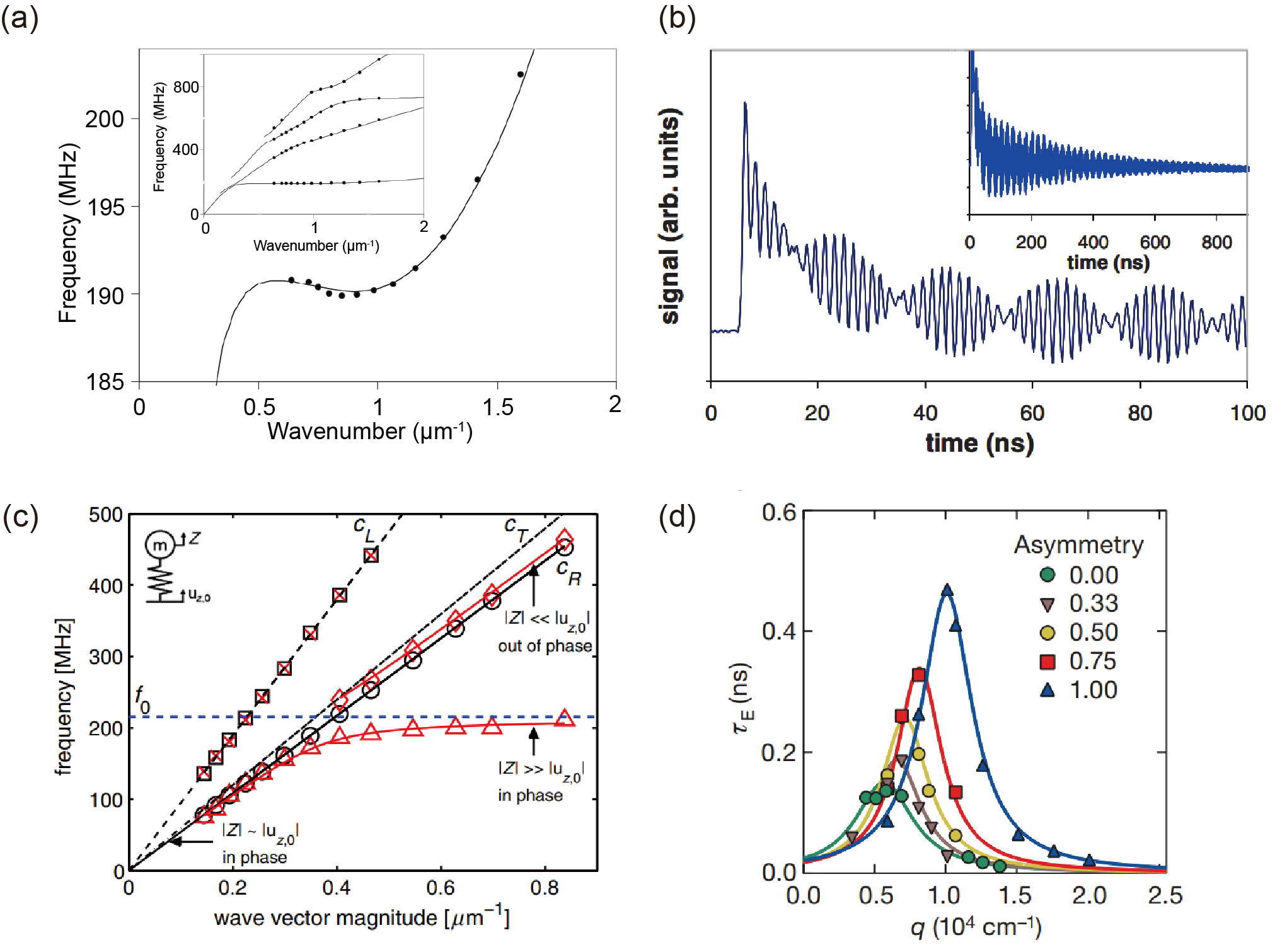}}
\caption{(a) Measured (solid circles) and calculated (line) dispersion relation of the lowest frequency surface acoustic mode in a silicon thin film (Inset: dispersion curves of the four lowest frequency modes)~\cite{maznev_apl_2009}. Reprinted from A. A. Maznev \textit{et al.}, \textit{Appl. Phys. Lett.} \textbf{95}, 011903 (2009), with the permission of AIP Publishing. (b) TGS signal from a micro-patterned 1D phononic structure, showing acoustic oscillations associated with non-leaky, long-lived surface acoustic modes. (Inset: the same signal over a longer time scale)~\cite{maznev_japplphys_2009}. Reprinted from A. A. Maznev \textit{et al.}, \textit{J. Appl. Phys.} \textbf{105}, 123530 (2009), with the permission of AIP Publishing. (c) Acoustic dispersion of a 2D granular crystal, where red and black markers represent the measured frequencies with and without silica microspheres, respectively. A horizontal dotted line indicates the resonance frequency arising from microsphere-substrate adhesion~\cite{boechler2013interaction}. Reprinted figure with permission from N. Boechler \textit{et al.}, Interaction of a Contact Resonance of Microspheres with Surface Acoustic Waves, \textit{Phys. Rev. Lett.} \textbf{111}, 036103 (2013). Copyright 2013 by the American Physical Society. (d) Measured spin helix lifetimes versus wavevector in GaAs/AlGaAs quantum wells, showing enhanced lifetime peaks to higher wavevectors with increasing doping asymmetry~\cite{koralek2009emergence}. Reproduced with permission from J. D. Koralek \textit{et al.}, Emergence of the persistent spin helix in semiconductor quantum wells, \textit{Nature} \textbf{458}, 7238 (2009). Copyright 2009 Springer Nature.}
\label{fig:ISBS_app}
\end{figure}

\section{Applications of ISS}
In impulsive stimulated scattering processes, surface acoustic waves (SAWs) are launched by a density grating from electrostriction or a thermal grating from optical absorption, depending on the nature of the medium. The wavelength of the excited SAW is determined by the optical grating period, enabling non-contact measurements of SAWs and guided modes in thin films, multilayers, liquids, and interfaces~\cite{maznev_jap_2009, kim_aplmat_2017, maznev_japplphys_2009, simmonds_aipadv_2024, yang_jchemphys_1995, yan_jchemphys_1988, paolucci_jchemphys_2000}. By measuring dispersion and velocities of surface acoustic modes, elastic moduli, film thickness, and depth profiling of structural defects can be obtained~\cite{rogers2000optical, steigerwald_apl_2009, mechri_apl_2009}. Figure~\ref{fig:ISBS_app}(a) shows experimental results on stacked crystalline Si with polycrystalline Cu-Ta thin film layers, where the lowest acoustic mode was observed to exhibit a negative group velocity at a frequency near 190 MHz, analogous to Lamb modes in free plates~\cite{maznev_apl_2009}. ISS can also provide access to interfacial acoustics; for instance, co-propagating surface modes and a short-lived, high-velocity interfacial mode were observed at TiO$_2$-water interfaces~\cite{kasinski_jcp_1989}, whereas at solid and glass-forming liquid interfaces, it has been applied to capture structural relaxation dynamics during the glass transition~\cite{glorieux_jap_2006}. Another application of ISS includes probing the quality of adhesion and mechanical properties of coatings, such as MnO$_2$ films, where SAW velocity dispersion and Scholte wave velocities were measured to estimate Young’s modulus and porosity of coatings~\cite{sermeus_jap_2014}.

ISS has also been employed to measure attenuation and dispersion relations of SAWs in phononic systems. In Fig.~\ref{fig:ISBS_app}(b), a micro-patterned SiO$_2$-Si structure exhibits TGS signal from spatially localized, non-leaky surface acoustic modes with lifetimes of hundreds of nanoseconds~\cite{maznev_japplphys_2009}. On the other hand, grooved silica substrate showed additional band gaps and leaky SAW modes~\cite{malfanti_jmechphyssol_2011}. In locally resonant acoustic metamaterials, hybridized flexural Lamb waves~\cite{khanolkar_apl_2015} and avoided crossings between contact resonances and Rayleigh modes, shown in Fig.~\ref{fig:ISBS_app}(c), were observed~\cite{boechler2013interaction, eliason_apl_2016, vega-flick_prb_2017}. More recently, ISS has moved from point spectroscopy to Brillouin microscopy of micromechanics in transparent and soft materials \cite{krug_optexp_2022, li_photonres_2024, krug2019impulsive, Jiarui_2022_Optics_ISBSapp1}. Recent work has demonstrated rapid elastography with sub-millisecond acquisition time, enabling 2D Brillouin shift mapping of PDMS–methanol mixture with reduced photodamage~\cite{li_photonres_2024}.

By adjusting pump beams to be cross-polarized, transient spin grating spectroscopy can be realized, which is used to probe spin dynamics of electrons and excitons, and valley polarization dynamics across various semiconductors and magnetic materials, including low-dimensional systems such as quantum wells and quantum dots~\cite{wang_natcommun_2013, walser_natphys_2012, Yang_natphys_2012, carter_prl_2006, yang_prl_2012, Mahmood_2018_NanoLetters_ISBSapp2,ishiguro_apl_2007, watanuki_apl_2005, scholes_prb_2006, kim_jphyschemb_2006, huxter_chemphyslett_2010, crisp_nanolett_2013}. Early seminal work using transient spin grating demonstrated that electron drift mobility and diffusion in GaAs/AlGaAs multiple quantum wells can be measured via exploiting 90\degree~polarization-rotation of a linearly polarized probe beam diffracted from spin grating, which isolates electron spin transport from ambipolar charge transport~\cite{cameron_prl_1996}. Wavevector–resolved spin grating subsequently revealed non-diffusive spin-polarization transport in semiconductor quantum wells, with the maximum spin decay time at a finite wavevector that contradicts predictions from simple diffusion theory due to spin-orbit coupling~\cite{weber_prl_2007}. In addition, spin Coulomb drag effect in a two-dimensional electron gas was demonstrated using spin gratings, where electron–electron scattering transfers momentum between up- and down-spin channels, thereby suppressing macroscopic spin transport~\cite{weber_nature_2005}. As shown in Fig.~\ref{fig:ISBS_app}(d), spin gratings have been used to probe the tunability of spin polarization lifetimes; by varying the level of doping asymmetry to manipulate the strength of Rashba spin-orbit interaction, it was demonstrated that the peak lifetimes of persistent spin helix modes in GaAs/AlGaAs quantum wells can be progressively shifted along wavevector~\cite{koralek2009emergence}. More recently, TGS has been extensively used to probe valley polarization dynamics in low-dimensional materials lacking inversion symmetry~\cite{Mahmood_2018_NanoLetters_ISBSapp2, wang_apl_2019, kuhn_laserphot_2020}, including the observation of fluence-dependent valley decay rate in monolayer MoSe$_2$, where strong exciton-exciton interactions lead to fast valley depolarization lifetimes of a few picoseconds~\cite{Mahmood_2018_NanoLetters_ISBSapp2}. Together, these studies highlight the functionality of TGS for probing spin and valley transport across systems from bulk crystals to emerging quantum materials, providing insights into spin-orbit coupling and valley excitons that are essential for developing spintronic, valleytronic, and optoelectronic devices. 

\begin{figure}
\centering
{\includegraphics[width=0.6\textwidth]{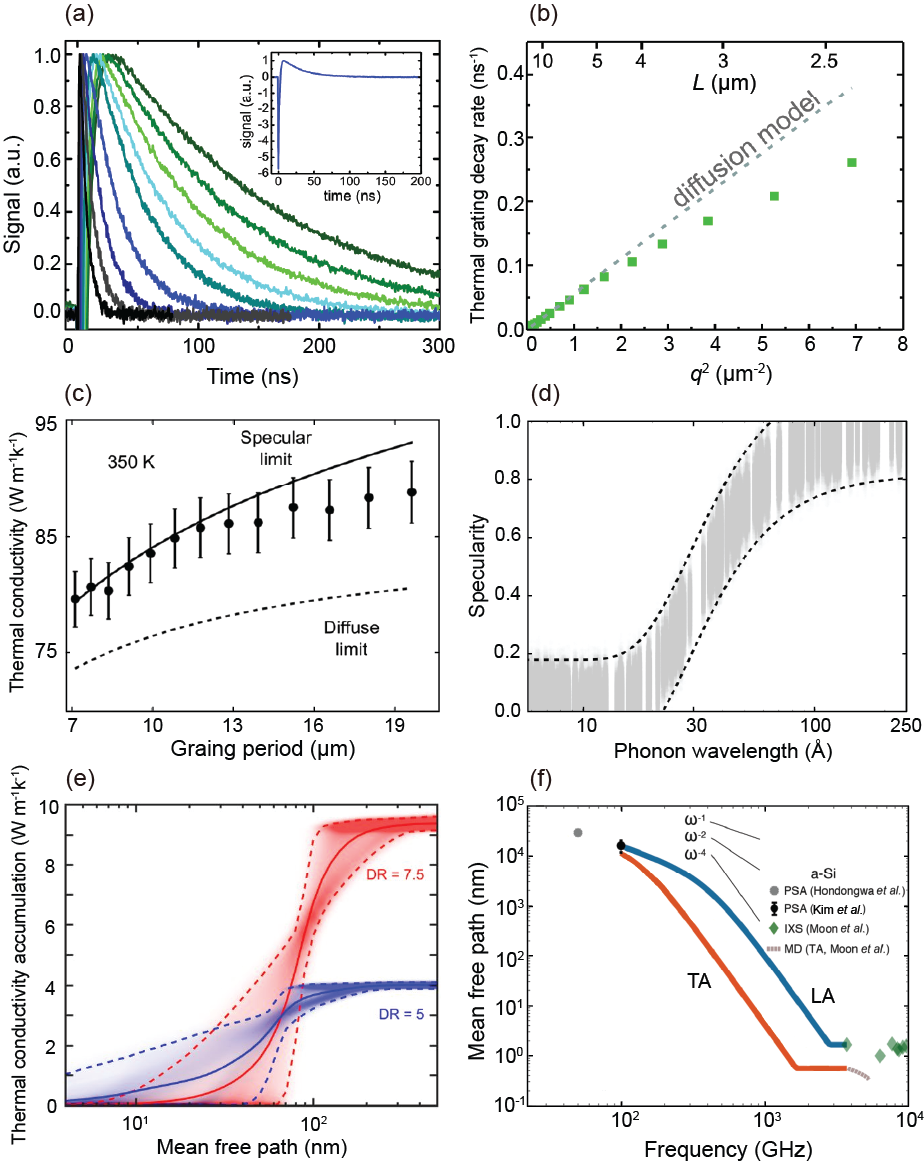}}
\caption{(a) Thermal decay signals of silicon at several transient grating periods. The inset shows the complete wave form for a representative grating period of 7.5 $\mu$m. The thermal signal decays slower as the grating period increases. Reprinted figure with permission from Johnson \textit{et al.}, \textit{Phys. Rev. Lett.}, \textbf{110}, 025901 2013. Copyright 2013 by the American Physical Society~\cite{Johnson_2013_PRL_TGsignals}. (b) Thermal decay rate versus grating wavevector squared, $q^2$~\cite{Johnson_2013_PRL_TGsignals}. As grating period decreases, the thermal decay rate deviates from what is predicted from heat diffusion theory. (c) Measured thermal conductivity versus grating period at 350 K in silicon thin film. As grating period decreases, the measured thermal conductivity approaches specular limit~\cite{Navaneetha_2018_PRX_specularity}. (d) Reconstructed phonon specularity versus phonon wavelength from grating period dependent thermal conductivity measurements including (c). As wavelength of phonon gets larger, specularity increases~\cite{Navaneetha_2018_PRX_specularity}. (e) Reconstructed thermal conductivity accumulation of partially ordered polyethylene thin films~\cite{Andrew_2019_PNAS_DrawRatio}. (f) Reconstructed frequency dependent MFPs of thermal acoustic excitations in aSi. Reprinted figure with permission from Kim \textit{et al.}, \textit{Phys. Rev. Materials}, \textbf{5}, 065602 2021. Copyright 2021 by the American Physical Society~\cite{Kim_2021_PRM_aSi_TG}.}
\label{fig:ISTS_app}
\end{figure}

To extract thermal properties, thermal signals from ISTS measurements, also known as transient thermal grating, have been analyzed. For materials where the optical penetration depth exceeds the sample thickness, the heat transfer can be treated as pure one-dimensional. As a result, the thermal gratings with period $L$ exponentially decay with an average decay time constant $\tau = 1/ (\alpha q^2)$ in which $\alpha$ is in-plane thermal diffusivity and $q=2\pi/L$ is the grating wavevector. This relation suggests that determining the thermal decay time constant at a given grating period enables measurement of the thermal diffusivity from which one can also calculate the thermal conductivity using the heat capacity. Figure~\ref{fig:ISTS_app}(a) shows representative ISTS thermal decay signals measured in silicon thin film (thickness: 400 nm)~\cite{Johnson_2013_PRL_TGsignals} at various grating periods ranging from 3.2 to 18 $\mu$m. The signals consist of two components: an initial rise that quickly decays after the photoexcitation, followed by slower thermal decay with a time constant of few tens of nanoseconds. The first signal component is due to ambipolar diffusion, while the second component corresponds to the thermal decay signal. For silicon, the ambipolar diffusion coefficient is larger by an order of magnitude, compared with the thermal diffusivity ~\cite{Li_PRB_1997, Eichler_2013_textbook}. Therefore, electronic and thermal relaxation dynamics can be distinctly resolved in the time domain. The decay time increases with the grating period and the decay rate $\gamma = 1/\tau$ is related to $q^2$ by a linear trend at sufficiently large grating period as shown in Fig.~\ref{fig:ISTS_app}(b). Here, the corresponding thermal conductivity at large grating period is the bulk value. However, as the grating period decreases, $\gamma$ versus $q^2$ deviates from the linear trend and the actual thermal decay becomes slower compared to the prediction from the heat diffusion theory. It has been established that the deviation is due to ballistic transport of thermal acoustic vibrations over the grating period.

At the onset of ballistic phonon transport, the thermal conductivity contribution from heat-carrying, thermal phonons with MFP ($\Lambda$) longer than the thermal length starts to decrease~\cite{Maznev_PRB_2011_supp}. The corresponding grating period dependent thermal conductivity is expressed as~\cite{Minnich_PRL_2012}
\begin{equation}
\kappa (q) = \int_{0}^{\infty} S(x) f(\Lambda_{\omega}) \, d\Lambda_{\omega} = \int_{0}^{\infty} q K(x)F(\Lambda_{\omega}) \, d\Lambda_{\omega}
\label{eqn:kTG_MFPdomain}
\end{equation}
where $x = q \Lambda_\omega$, and $K(x)=-dS/dx$ with $S(x)$ being the heat suppression function that accounts for effective heat flux, which is reduced compared to Fourier law predicted heat flux due to quasi-ballistic transport \cite{Minnich_PRL_2012}. The cumulative thermal conductivity, $F(\Lambda_{\omega}) (= \int_{0}^{\Lambda_\omega} f(\Lambda) d\Lambda)$, provides insight into microscopic heat conduction properties by enabling to examine the contributions of MFPs of heat carriers to the thermal conductivity. By changing the integration variable from MFP to frequency~\cite{Yang_PRB_2013}, one can also express Eq.~\ref{eqn:kTG_MFPdomain} as
\begin{equation}
\kappa (q) = \sum_\mathbf{p} \int_{0}^{\infty} S(x_{\mathbf{p}}) \left[ C_{\mathbf{p}}(\omega) v_\mathbf{p} (\omega) \Lambda_\mathbf{p}(\omega) \right] d \omega
\label{eqn:kTG_omegadomain}
\end{equation}
where $\mathbf{p}$ indexes the polarization and $C_\mathbf{p}$ and $v_\mathbf{p}$ are the specific heat and group velocity of thermal energy carriers. In Eq.~\ref{eqn:kTG_omegadomain}, $\Lambda_\mathbf{p}(\omega)$ is MFPs versus frequency, representing scattering mechanisms that depend on the frequency. In TGS, it has been established that the cumulative thermal conductivity and the frequency versus MFPs can be reconstructed by solving Eqs.~\ref{eqn:kTG_MFPdomain} and~\ref{eqn:kTG_omegadomain}, and that ill-posed problems of Eqs.~\ref{eqn:kTG_MFPdomain} and~\ref{eqn:kTG_omegadomain} can be solved using numerical optimizations or statistical methods~\cite{Minnich_PRL_2012, Navaneetha_2018_PRX_specularity, Andrew_2019_PNAS_DrawRatio}.

The measurements of the grating period dependent thermal conductivity in macroscopic thin films have provided insights into the understanding of microscopic transport properties governing heat conduction along the in-plane direction. It should be noted that heat suppression functions have been obtained by solving the phonon Boltzmann transport equation (BTE) within the framework of relaxation time approximation (RTA) or Callaway model to describe lattice thermal conductivity~\cite{Shiga_PRB_2012, Skelton_PRB_2014, Minnich_PRB_2015}. However, these approximations show limitations in describing phonon transport, especially in high thermal conductivity materials and in the non-diffusive heat flow regime where momentum-conserving phonon scattering processes dominate~\cite{Lindsay_PRB_2013, Jinlong_PRB_2014, Malviya_PRB_2023, Malviya_J.Phys.Condens.Matter_2025}. Therefore, symmetry properties of the linearized scattering operator and low-rank representations of the solution have been suggested to accurately solve the BTE with reduced computational cost when the RTA and Callaway assumptions are no longer valid~\cite{Chaput_PRL_2013, Hua_PRB_2020, Navaneetha_arXiv_2025}.

While it has been recognized that atomic scale imperfections such as surface roughness strongly affects phonon boundary scattering, hence the effective thermal resistance, experimentally resolving the boundary conditions at which phonon specular reflections occur has been challenging \cite{Chen_NatRevPhys_2021}. In estimating specularity parameter, numerical models such as the Ziman's specularity model~\cite{Ziman_2001} have failed to fully explain the specularity parameter and often yield contradictory results~\cite{Cuffe_PRL_2013, Hertzberg_NanoLett_2014, Jiang_PRB_2018} . Furthermore, prior studies on thermal transport in silicon nanowires have reported that the unphysical specularities, which are incompatible with Ziman's theory, yield the best agreement with the measured temperature-dependent thermal conductivity trend~\cite{Mingo_PRB_2003, Kazan_JAP_2010}. Above problems can be attributed to the fact that while the measured signal represents an ensemble average over the spot size, the Ziman's model depends on atomic-scale local properties such as surface roughness and impurities, which complicates the analysis. Overall, these imply that it is necessary to experimentally resolve the parameter as a function of phonon wavelength. Ravichandran \textit{et al.} have demonstrated a metrology to extract specular parameters from the ISTS measurements on crystalline silicon thin films, as shown in Fig.~\ref{fig:ISTS_app}(c)~\cite{Navaneetha_2018_PRX_specularity}. The experimentally resolved specularity is shown in Fig.~\ref{fig:ISTS_app}(d), in which the spectral dependence of phonon reflections at the boundary can be seen.

Additionally, ISTS measurements on thermally conductive polymers~\cite{Andrew_2019_PNAS_DrawRatio,Kim_Natcomm_2022} have revealed microscopic transport properties that lead to exceptional thermal conductivities. It has been shown that the thermal conductivities of structurally-engineered semi-crystalline polymers including polyethylene (PE) can achieve orders of magnitude increases, which can be as high as those in many condensed matters. While such thermal conductivity enhancements have been typically interpreted using effective medium models~\cite{takayanagi_JPSC_1964, choy_polymer_1977, Lu_JAP_2018}, debate has existed concerning the origin of the high thermal conductivity, particularly for highly oriented polymers~\cite{ronca_polymer_2017, Xu_Natcomm_2019}. Recent ISTS measurements applied to stretched PEs have provided insights into microscopic origins that lead to high thermal conductivities in these specimens. As shown in Fig.~\ref{fig:ISTS_app}(e), for partially oriented PEs, it has been shown that MFP of phonons that substantially contribute to the heat conduction is on the order of 100 nm, which exceeds the grain size of $\sim 60$ nm, indicating ballistic propagation of phonons across the grain boundaries~\cite{Andrew_2019_PNAS_DrawRatio}. Additionally, more recent measurements on highly oriented PEs have shown that thermal conductivity of macroscopic thin films of PE can be as high as $\sim 50$~\wmk, which was primarily attributed to thermal acoustic phonons that propagate over a few hundreds nanometers, limited by enlarged crystals in them~\cite{Kim_Natcomm_2022}.

ISTS experiments have also been applied to amorphous solids and experimentally probed the scattering of thermal vibrations that occurs distinctly in them, which was not captured using prior numerical studies~\cite{Kim_2021_PRM_aSi_TG}. An experimental approach that enables to reconstruct the frequency dependent MFPs along with MFP accumulation function by utilizing constraints were demonstrated and applied to  amorphous silicon (aSi). Results are shown in Fig~\ref{fig:ISTS_app}(f), in which power-law dependence of MFPs ($\Lambda \sim \omega^{-n}$) can be seen, indicating the characteristics of scattering that depends on the frequency. In particular, Rayleigh-type scattering as a primary scattering mechanism ($n=4$) was observed above 1 THz for aSi, which contradicts with predictions from prior normal mode analysis yet agrees with experimental results in other glassy substances such as vitreous silica.

\section{Outlook and Conclusion}
To make further developments in Brillouin scattering, recent efforts have focused on pushing the limits of the spatiotemporal resolution of the technique. The introduction of short wavelength light sources in TGS setups has enabled grating periods approaching sub-micrometer length scales, allowing researchers to probe non-diffusive dynamics in materials. Free electron lasers in the extreme ultraviolet (EUV) \cite{Bencivenga_2015_nature_EUV_TGS_1, Bencivenga_2019_scienceAdv_EUV_TGS_2, Miedaner_2024_scienceadv_EUV_TGS_3} and X-ray \cite{Rouxel_2021_naturePhotonics_XrayTG} bands have been implemented to further increase penetration depth and push grating period limits below nanometer scales. These grating periods can elucidate magnetic dynamics and unconventional heat transport at nanometer scales. In the time domain, fast electronic processes such as electron-hole recombination can be temporally resolved with attosecond-scale temporal resolution by implementing both sub-femtosecond excitation and detection processes~\cite{Fidler_2019_Natcom_atto1, Bermudez_2025_ACS_atto2}, providing valuable insights into fast electronic processes such as carrier thermalization and recombination processes. Furthermore, when combined with enhanced temporal and spatial resolution, spin gratings also have potential for studying electronic spin transport dynamics that can establish the theoretical background for spintronic device applications.

Brillouin scattering has also expanded into applications beyond materials science, including optical communications and biological imaging. In photonics, stimulated Brillouin processes have been investigated to exploit enhanced Brillouin gain in optical fibers and waveguides for applications in microwave photonics and signal processing~\cite{Kobyakov_advoptphoton_2010, wolff_pra_2015, botter_sciadv_2022}. This enables ultra-low-loss on-chip platforms that provide narrowband optical gain, demonstrating the feasibility of integrated Brillouin photonics~\cite{neijts_aplphot_2024, ye_aplphot_2025}. In the field of life sciences, Brillouin microscopy has emerged as one of the instrumental techniques for mapping mechanical properties of soft and liquid biological samples. A pioneering work reported the first in situ three-dimensional imaging of the intraocular lens of a mouse using confocal Brillouin microscopy~\cite{Scarcelli_2008_NatPhotonics_ISBSapp4}. Since then, Brillouin technique has been applied to diverse biological systems, including 3D imaging of human cornea both in vivo~\cite{Scarcelli_optexp_2012} and ex vivo~\cite{scarcelli_invopht_2013}, probing intracellular stress granule mechanics during liquid-to-solid phase transition~\cite{Antonacci_communbio_2018}, and mapping intracellular biomechanics within a whole living organism~\cite{Scarcelli_natmet_2015}. These advances highlight the possible translation of Brillouin microscopy from basic research to clinical diagnostics, particularly in ophthalmology~\cite{shao_scirep_2019, Kabakova_natrev_2024, Bouvet_natphot_2025}.

To conclude, Brillouin scattering spectroscopy remains an indispensable technique for probing low-frequency acoustic and spin waves in thin films and bulk materials. By providing access to these sub-THz excitations, it provides information on dispersions and lifetimes of these quasiparticles for understanding energy transport in novel materials. By integrating with other foundational inelastic scattering techniques, such as Raman spectroscopy, INS, and IXS, Brillouin scattering spectroscopy can bridge the gap in length and time scales of energy transport processes that were previously inaccessible. This provides a quantitative experimental platform to gain insight into nanoscale energy transport and conversion processes, with direct implications for guiding the design of thermoelectrics, thermal interface materials, and nanostructured systems for energy harvesting, thermal management, and microelectronics.

\section*{Acknowledgements}
This work was supported by the National Research Foundation of Korea (NRF) grant funded by the Korea government(MSIT)(No. RS-2023-00211070 and No. RS-2024-00403173).

\section*{Author contributions}
Hyemin Kim: Writing - original draft (equal); Writing - review \& editing (equal). Hyungseok Kim: Writing - original draft (equal); Writing - review \& editing (equal). Taeyong Kim: Conceptualization (lead); Funding acquisition (lead); Supervision (lead); Writing - original draft (supporting); Writing - review \& editing (equal).

\section*{Data availability}
Data sharing is not applicable to this article as no new data were created or analyzed in this study.

\section*{Competing interests}
The authors have no conflicts to disclose.

\bibliographystyle{unsrtnat}

\bibliography{InelasticReviewRef_rev}

\end{document}